\documentclass[seceq,preprint]{article}
\usepackage{graphicx}

\begin{document}

\markboth{    
Rosu, Reyes, Valencia 
}{            
Three-Step Master Equation
}

\centerline{ArXiv: math-ph/0402068 v2}

\bigskip

\centerline{
THE THREE-STEP MASTER EQUATION: CLASS OF PARAMETRIC STATIONARY SOLUTIONS
}

\bigskip

\centerline{       
Haret C. \textsc{Rosu\footnote{e-mail: hcr@ipicyt.edu.mx}}$^{a}$, Marco A. \textsc{Reyes} $^{b}$, F. \textsc{Valencia}$^{a}$%
}

\begin{center}
$^{a}$ Potosinian Institute of Science and Technology,\\
Apdo Postal 3-74 Tangamanga, 78231 San Luis Potos\'{\i}, Mexico\\

$^{b}$ Institute of Physics, University of Guanajuato, \\ Apdo Postal E-143, Le\'on, Gto, Mexico\\

\end{center}

\bigskip



{\small

We examine the three-step master equation 
from the standpoint of the general solution of the associated discrete Riccati equation. We report by this means stationary master solutions depending 
on a free constant parameter, denoted by $D$, that should be negative 
in order to assure the positivity of the solution. These solutions correspond to different discrete Markov processes characterized by
the value of $D$, which is related to specific renormalizations of the transition rates of the chain of states.}

\bigskip




In general, the three-step population master equation is used by physicists 
in many studies of diffusion processes of microscopic particles on one-dimensional lattices \cite{j}, but this simple discrete equation has extensive and interesting applications 
in other fields as well, most recently to Hubbell's neutral theory in ecology \cite{ban}. In the following, we shall use a population interpretation. It reads
\begin{equation} \label{M1}
\frac{dp_{n}}{dt}=d_{n+1}p_{n+1}- \sigma _{n}p_{n}+b_{n-1}p_{n-1}~, \qquad \qquad   \sigma _{n,}=b_{n}+d_{n}~,
\end{equation}
where $b_{n}$ is the transition rate for the birth-type jump $n\rightarrow n+1$  
and $d_{n}$ is the death-type rate for the backward jump $n\rightarrow n-1$,  
while
$p_{n}$ is the probability to have $n$ individuals at the instant $t$.
Employing the initial conditions $b_{-1}=d_{0}=0$,
the known stationary solution is, \cite{G}
\begin{equation} \label{M2}
P_{n}^{{\rm st}}= P_{0}
\Bigg(\prod _{j=0}^{n-1} \frac{b_{j}}{d_{j+1}}\Bigg)~,
\end{equation}
where $P_{0}$ is a scaling constant that through the probabilistic normalization
condition can be written as $P_{0}=\Bigg(1+\sum _{n=1}^{N}
\prod _{j=0}^{n-1} \frac{b_{j}}{d_{j+1}}\Bigg)^{-1}$ (see, \cite{gil}).

\medskip



\noindent
We proceed now to show that stationary solutions which are different of Eq.~(\ref{M2}) can be obtained 
that are based on the general solution of the discrete Riccati equation connected to the 
master equation. 
Indeed, performing the transformation  
\begin{equation}\label{M3}
y_{n-1}=\frac{P_{n-1}}{P_{n}}+\frac{1-\sigma _{n}}{b_{n-1}}~, \qquad n\neq 0~,  
\end{equation}
in Eq. (\ref{M1}) leads to the following discrete Riccati equation
\begin{equation}\label{M4}
y_{n}=b_{n-1}y_n y_{n-1} -      
\frac{b_{n-1}}{b_n}(1-\sigma _{n+1})y_{n-1}
+d_{n+1}+\frac{1-\sigma _{n+1}}{b_n}~,
\end{equation}
with the particular solution
\begin{equation}\label{M5}
y_{n}^{0}=\frac{1-b_{n+1}}{b_{n}}~.
\end{equation}
However, it is easy to check that one can write a more general solution of Eq.~(\ref{M4}) as follows   
\begin{equation}\label{M7}
y_{n}^{1}=y_{n}^{0}+\frac{f_{n}}
{D-\sum _{k=0}^{n}\frac{f_{k}b_{k+1}}{d_{k+2}}
}~,\qquad f_n=\prod _{i=0}^{n}\frac{b_id_{i+2}}{b^{2}_{i+1}}~,
\end{equation}
where $D$ is a real constant.

Using simple discrete algebra, one can obtain the recurrence relationship
\begin{equation}\label{Rec}
P_{n+1}=P_n\left(y_n+\frac{\sigma _{n+1}-1}{b_n}\right)^{-1}
\end{equation}
leading to stationary solutions of the following form
\begin{equation}\label{M8}
P_{n}(D)=\tilde{P_0}\prod _{i=0}^{n-1}\frac{b_i}{d_{i+1}}
\Bigg(1+\frac{f_i b_i/d_{i+1}}{|D|+\sum _{j=0}^{i}\frac{f_jb_{j+1}}{d_{j+2}}-\frac{f_i b_i}{d_{i+1}}}\Bigg)
\end{equation}
where the normalization constant reads
\begin{equation} \label{M8b}
\tilde{P_0}=\Bigg[1+  \sum _{n=1}^{N} \prod _{i=0}^{n-1} \frac{b_{i}}{d_{i+1}}    \Bigg(1+\frac{f_ib_i/d_{i+1}}{|D|+\sum _{j=0}^{i}\frac{f_jb_{j+1}}{d_{j+2}}-\frac{f_ib_i}{d_{i+1}}}\Bigg)\Bigg]^{-1}~.
\end{equation}


\noindent
Examining Eq.~(\ref{M8}) with normalization (\ref{M8b}), we first notice that for $D\rightarrow -\infty$ we recover the known case of stationary 
master solution with the common normalization. 
In addition, we notice that the factor 
$$
\Bigg(1+\frac{f_i b_i/d_{i+1}}{|D|+\sum _{j=0}^{i}\frac{f_jb_{j+1}}{d_{j+2}}-\frac{f_i b_i}{d_{i+1}}}\Bigg)
$$
looks like a renormalization factor for the transition rates of the original stationary Markov process. 
A reasonable interpretation of $D$ depends on the specific application and in 
general is related to initial conditions, boundary conditions, or external applied fields.
In addition, for a physical solution one requires positivity implying
$$
\frac{b_i}{d_{i+1}}\geq \frac{b_{i-1}}{d_{i}}~.
$$
This is a strong condition and in particular cases it could be relaxed.


\medskip

\noindent
{\bf a}). The most trivial case is $b_k=d_k={\rm const} < 1,\, k=0,..., n$.
This implies $P_{k}^{st}=P_0$; the Riccati solution is $y_k=b^{-1}-1$.
The $D$-dependent solution will be 
\begin{equation}\label{pc1}
P_k=\tilde{P}\left(1+\frac{k}{D}\right)
\end{equation}
and with the normalization explicitly calculated
\begin{equation}\label{pc2}
P_k=\frac{1}{n+1}\left(\frac{1+\frac{k}{D}}{1+\frac{n}{2D}}\right)~.
\end{equation}

A plot of $P_k$ for various values of the parameter $D$ is shown in Fig.~(1).

\medskip

\noindent
{\bf b}). For the asymmetric case we take $b_n=\frac{1}{2}(1+\epsilon), \, d_n=\frac{1}{2}(1-\epsilon), \,q=\frac{1-\epsilon}{1+\epsilon}<1,\, n=0,...N$,
and the parametric solution reads
\begin{equation}\label{pc3}
P_n=\tilde{P}_0q^{-n}\prod _{i=0}^{n-1}\frac{(1-q)(D+q^i)+1-q^i}{(1-q)D +1-q^i}~,
\end{equation}
and the normalization constant can be easily written down from Eq.~(\ref{M8b}). Plots for this case are displayed in Fig.~(2).

\medskip

\noindent
{\bf c}). Various other cases are presented in Figures~(3) - (5) for exponential parametrizations of the jump rates. 

\medskip



\noindent
In summary, we report one-parameter stationary solutions of the three-step master equation
that are based on the corresponding discrete Riccati general solution.
In the continuous case, the mathematical method we employ here corresponds to
Mielnik's procedure in supersymmetric quantum mechanics \cite{miel}. The parameter of these solutions
could be fixed in applications by initial/boundary conditions or through external perturbations of the underlying birth-death Markov process.  

It is well known that the stationary solution of the master equation of a discrete Markov process is uniquely defined if the 
process contains only one class of ergodic states. In this case, the stationary solution does not depend on the initial condition.
The discrete Riccati mathematical procedure leads to modified transition rates and consequently to different stationary solutions that belongs to different 
master equations.
Modifying the rates at the ends of the chain of states corresponds to a probability current flow through the system, i.e., to a 
driven system.
Thus, the physical interpretation of this class of parametric master solutions is that they are a specific type of current-carrying solutions that are important 
non-equilibrium steady states in many mesoscopic and macroscopic systems, such as Becker-D\"oring nucleation processes \cite{wattis04},  
or superconductivity, where as stated by Geller \cite{geller}, ``it is now understood that supercurrent-carrying states are in fact, metastable {\em non-equilibrium 
states} ... with an extremely long lifetime". The latter states are essential for the tunable supercurrent of Josephson junction technology \cite{baselmans}.  

If, for example, we place us in a population (ecology) context, the difference with respect to the original master equation is already at the level of the $d_0$ rate.
For a finite $D$ the rate $d_0$ is not zero. Thus, for positive $d_0$, one can interpret this rate as an initial immigration rate, while for
negative values as an initial emigration rate.  



\begin{figure}[htb]
\centerline{
\includegraphics[scale=0.9]{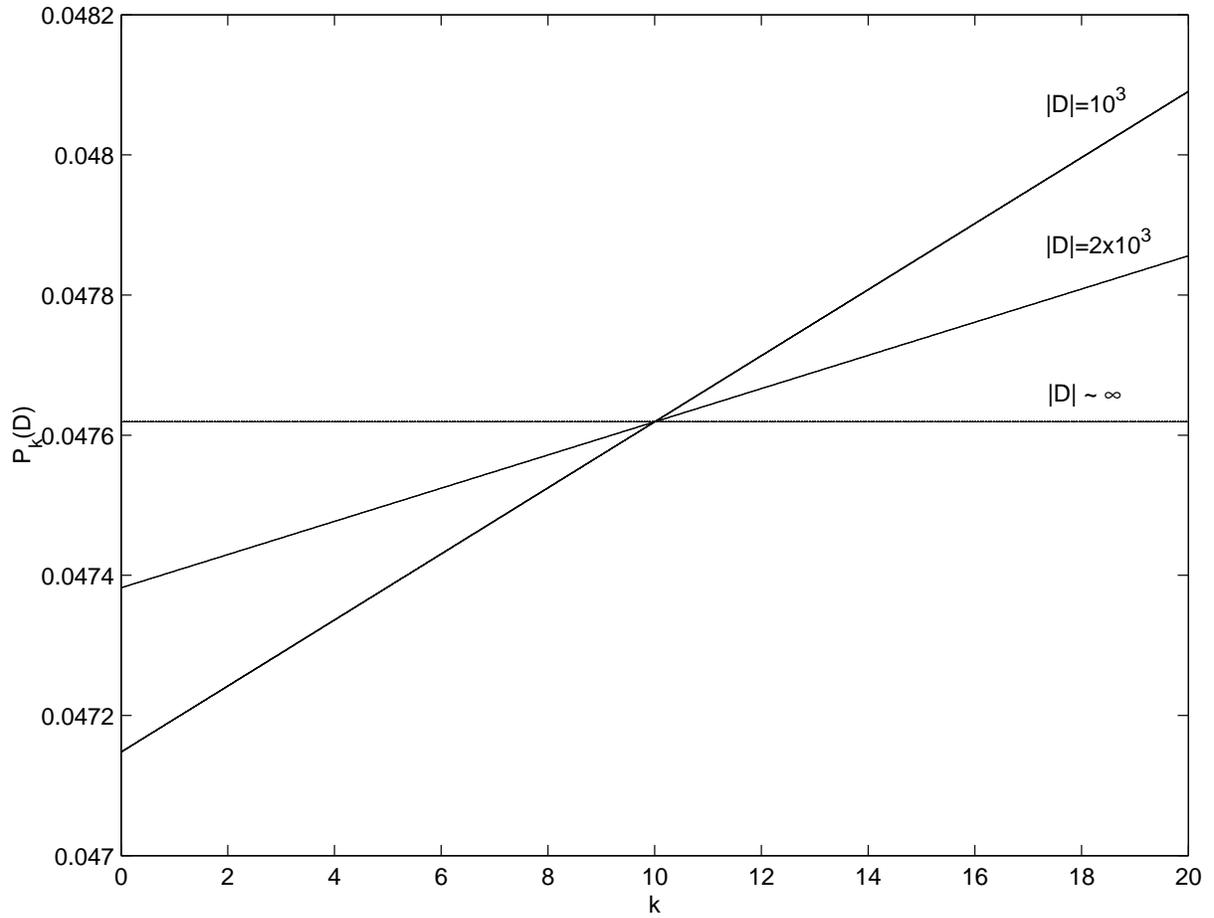}}
\caption{$P_k(-1000)$, $P_k(-2000)$ and $P_k(-\infty)$ for $b_i=d_i={\rm const}<1$ and $n=20$.
} \label{f1M}  
\end{figure}

\begin{figure}[htb]
\centerline{
\includegraphics[scale=0.7]{graph_5.eps}}
\caption{$P_n(-4)$ - black straight line; $P_n(-40)$ - blue dotted line; $P_n(-\infty)$ - red-dotted line,
all of them for $\epsilon=0.02$.
} \label{f2M}  
\end{figure}

\begin{figure}[htb]
\centerline{
\includegraphics[scale=0.7]{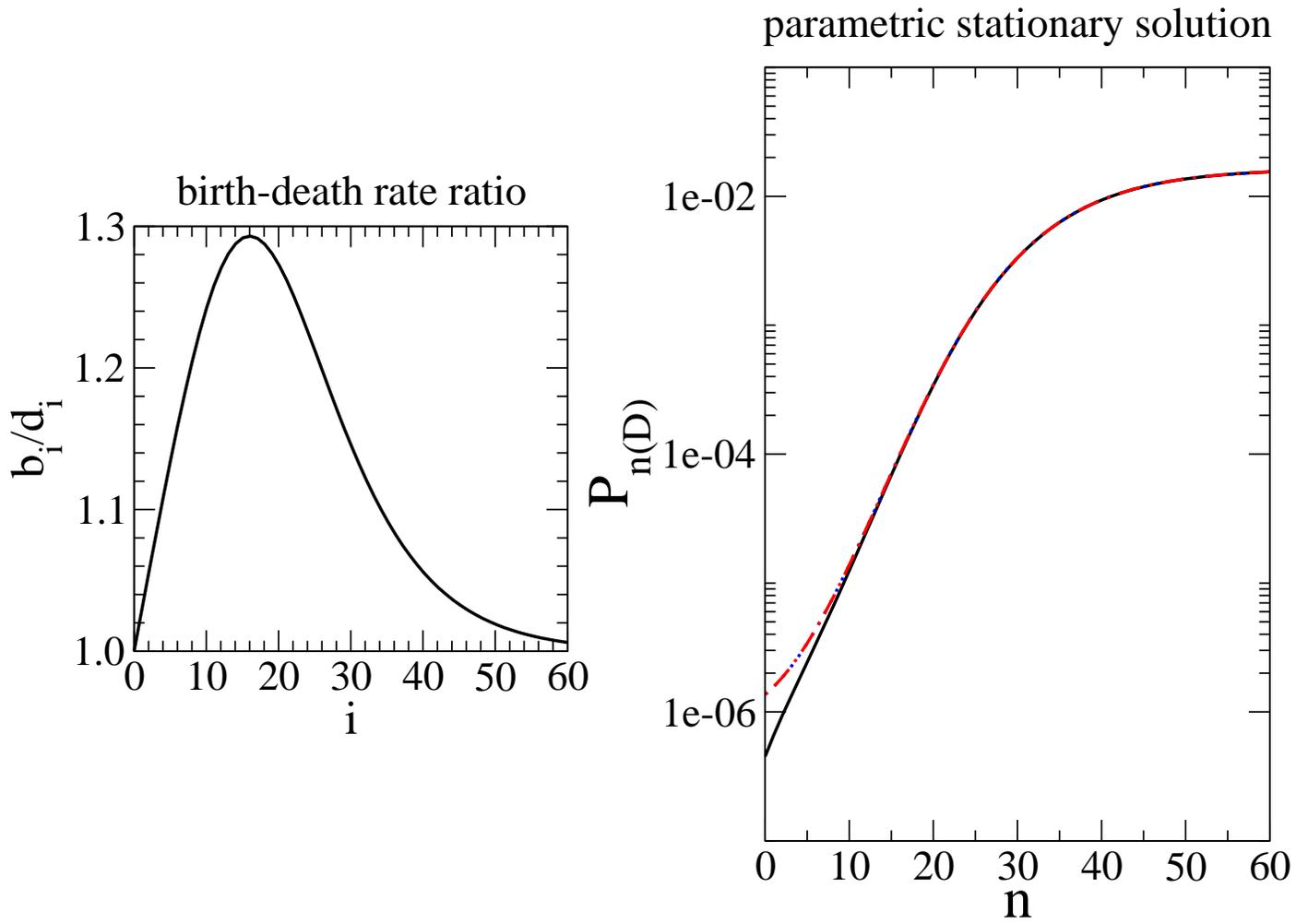}}
\caption{Black straight line - $P_n(-4)$; blue-dotted line is $P_n(-4000)$; red-dotted-dashed line is $P_n(-\infty)$- for $b_i=0.1+ \exp[-0.12\,i]$ and $d_i=0.1+ \exp[-0.15\,i]$.
} \label{f3M}  
\end{figure}

\begin{figure}[htb]
\centerline{
\includegraphics[scale=0.7]{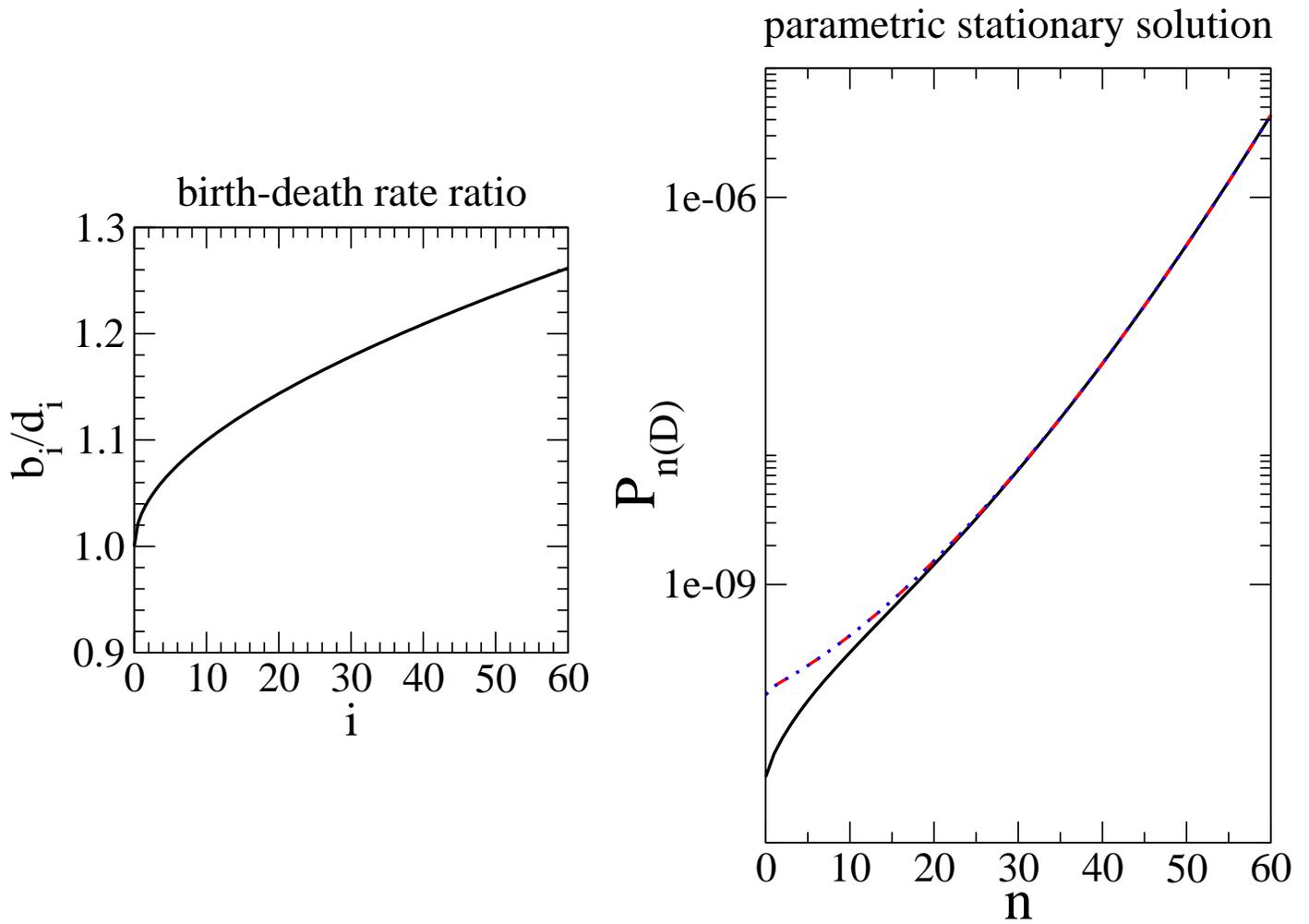}}
\caption{$P_n(-4)$ - black straight line; $P_n(-4000)$ - blue dotted line; $P_n(-\infty)$ - red-dotted line, for $b_i= \exp[-0.12\,i^{1/2}]$ and $d_i=\exp[-0.15\,i^{1/2}]$.
} \label{f4M}  
\end{figure}

\begin{figure}[htb]
\centerline{
\includegraphics[scale=0.7]{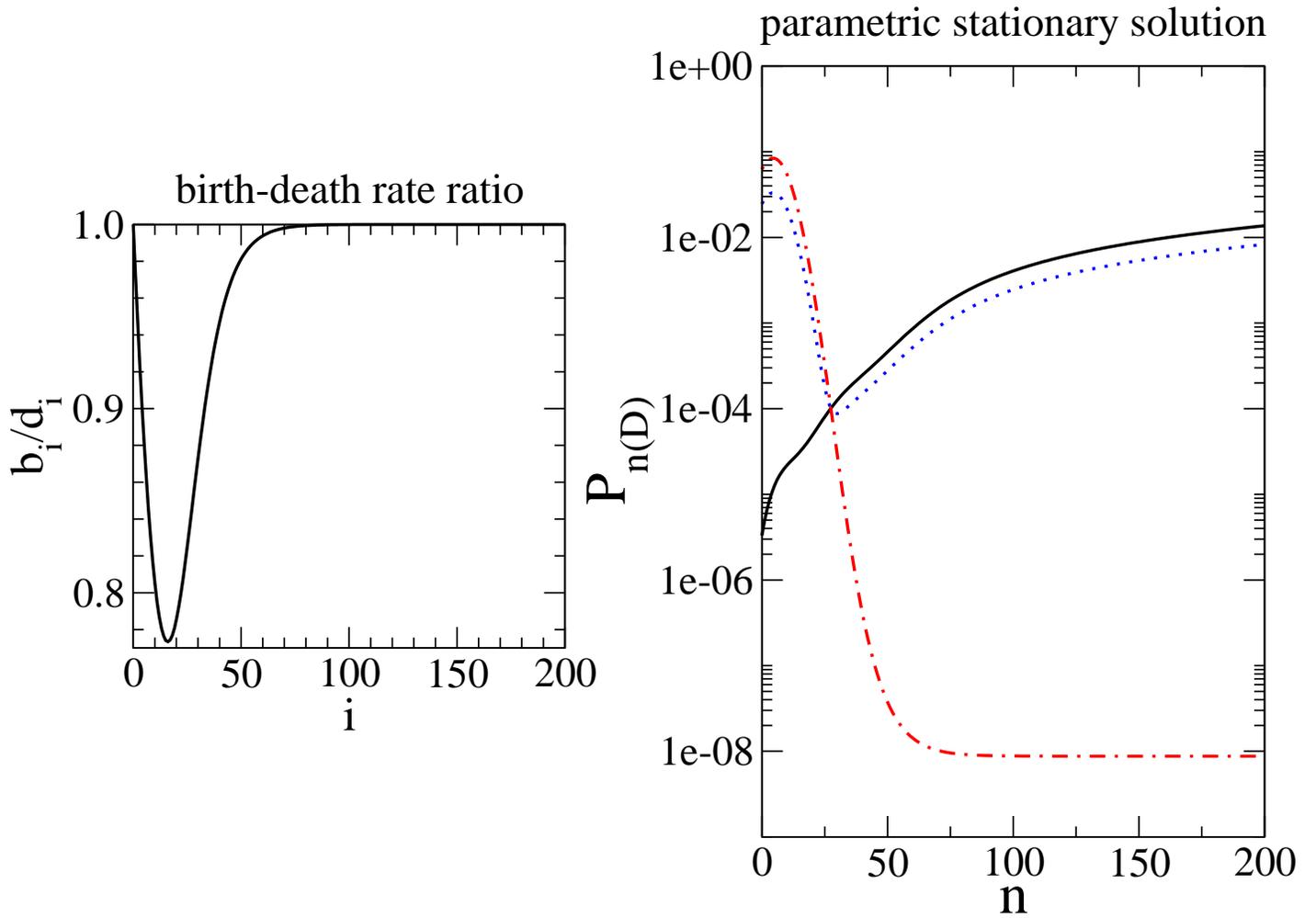}}
\caption{$P_n(-4)$ - black straight line; $P_n(-4000)$ - blue dotted line; $P_n(-\infty)$ - red-dotted line,
for $b_i= 0.01+\exp[-0.15\,i]$ and $d_i=0.01+\exp[-0.12\,i]$.
} \label{f5M}  
\end{figure}


%

\end{document}